\documentclass[aps,showpacs,nofootinbib,superscriptaddress,twocolumn]{revtex4}

\usepackage{graphicx}
\usepackage{amssymb}
\usepackage{amsmath}
\usepackage{bm}
\usepackage{slashed}
\usepackage{datetime}

\begin{document}
\allowdisplaybreaks[2]

\title{
Constraining quark angular momentum through semi-inclusive
  measurements}

\author{Alessandro Bacchetta}
\email{alessandro.bacchetta@unipv.it}
\affiliation{Dipartimento di Fisica Nucleare e Teorica, Universit\`a di Pavia, and}
\affiliation{INFN Sezione di Pavia, via Bassi 6, I-27100 Pavia, Italy}

\author{Marco Radici}
\email{marco.radici@pv.infn.it}
\affiliation{INFN Sezione di Pavia, via Bassi 6, I-27100 Pavia, Italy}

\begin{abstract}
The determination of quark angular momentum requires the knowledge of the
generalized parton distribution $E$ in the forward limit. We assume a connection 
between this function and the Sivers transverse-momentum distribution, based 
on model calculations and theoretical considerations. Using this assumption, we 
show that it is possible to fit at the same time nucleon magnetic moments and 
semi-inclusive single-spin asymmetries. This imposes additional constraints on
the Sivers function and opens a plausible way to quantifying quark angular
momentum. 
\end{abstract}

\maketitle

Nucleons are spin-1/2 composite particles made by partons (i.e., quarks and gluons).  
Determining how much of the nucleons' spin is carried by each parton is a critical endeavour towards an understanding of the microscopic structure of matter. In this work, we propose a way to constrain the longitudinal angular momentum $J^a$ of a (anti)quark with flavor $a$. To do this, we adopt an assumption, motivated by model calculations and theoretical considerations, that connects $J^a$ to the Sivers transverse-momentum distribution (TMD) measured in semi-inclusive deep-inelastic scattering (SIDIS)~\cite{Burkardt:2002ks}. The Sivers function 
$f_{1T}^{\perp a}$~\cite{Sivers:1990cc} is related to the distortion of the momentum distribution of an unpolarized parton $a$ when the parent nucleon is transversely polarized. We show that this assumption of relating $J^a$ to $f_{1T}^{\perp a}$ is compatible with existing data, and we derive estimates of $J^a$.

The total longitudinal angular momentum of a parton $a$ (with $a = q, \bar{q}$) at some scale $Q^2$ can be computed as a specific moment of generalized parton distribution functions (GPD)~\cite{Ji:1997ek}
\begin{equation}
J^{a} (Q^2)= \frac{1}{2}\, \int_0^1 dx \, x\, \Bigl(H^{a}(x,0,0; Q^2)+E^{a}(x,0,0; Q^2) \Bigr).
\label{e:Jdef}
\end{equation} 
The GPD $H^a(x,0,0; Q^2)$ corresponds to the familiar collinear parton distribution function (PDF) 
$f_1^{a}(x; Q^2)$, which gives the probability of finding at the scale $Q^2$ a parton with flavor $a$ and fraction $x$ of the (longitudinal) momentum of the parent nucleon. The forward limit of the GPD $E^a$ does not correspond to any collinear PDF~\cite{Diehl:2003ny}. It is possible to probe the function $E^a$ in experiments, but never in the forward limit (see, e.g., \cite{Kumericki:2007sa}). Assumptions are eventually necessary to constrain $E^a(x,0,0; Q^2)$. This makes the estimate of  $J^a$ particularly challenging. The only model-independent constraint is the scale-independent sum rule  
\begin{equation}
\sum_q \  e_{q_v} \int_0^1 dx \, E^{q_v}(x,0,0) = \kappa ,
\label{e:sumrule}
\end{equation} 
where $E^{q_v}=E^q-E^{\bar{q}}$ and $\kappa$ denotes the anomalous magnetic moment of the parent nucleon.

Inspired by results of spectator
models~\cite{Burkardt:2003je,Lu:2006kt,Meissner:2007rx,Bacchetta:2008af,Bacchetta:2010si}  and theoretical considerations~\cite{Burkardt:2002ks}, we propose the following simple relation at a specific scale $Q_L$,  
 \begin{equation}
 f_{1T}^{\perp (0) a}(x; Q_L^2) = -L(x)\, E^a (x,0,0; Q_L^2) ,
\label{e:EtoSivers0}
\end{equation}  
where we define the $n$-th moment of a TMD with respect to its transverse momentum $k_{\perp}$ as
\begin{equation}
f_{1T}^{\perp(n) a} (x; Q^2) = \int d^2 k_{\perp} \left( \frac{k_{\perp}^2}{2M^2} \right)^n \, 
f_{1T}^{\perp a} (x, k_{\perp}^2; Q^2)  , 
\label{e:n-mom}
\end{equation}
and $M$ is the nucleon mass. 

In Eq.~\eqref{e:EtoSivers0}, $L(x)$ is a flavor-indepedent function, representing the effect of the QCD interaction of the outgoing quark with the rest of the nucleon. The name ``lensing function'' has been proposed by Burkardt to denote $L(x)$~\cite{Burkardt:2003uw}. Computations of the lensing function beyond the single-gluon approximation have been proposed in 
Ref.~\cite{Gamberg:2009uk}.  It is likely that in more complex models the above relation is 
not preserved, at least not as a simple product of $x$-dependent functions~\cite{Meissner:2007rx}. Nevertheless, it is useful and interesting to speculate on the consequences of this simple assumption. As a more refined picture of TMD and GPD emerges, it will be possible to improve the reliability of this assumption or eventually discard it. The present attempt should be considered as a ``proof of concept'' for further studies in this direction.

The advantage of adopting the Ansatz of Eq.~\eqref{e:EtoSivers0} is twofold: first, it allows us to use the value of the anomalous magnetic moment to constrain the integral of the valence Sivers function; second, it allows us to obtain flavor-decomposed information on the $x$-dependence of  the GPD $E$ and ultimately on the quark total angular momentum. This is an enticing example of
how assuming model-inspired connections between GPD and TMD can lead to powerful outcomes.

The Sivers function has been extracted from SIDIS measurements by three 
groups~\cite{Vogelsang:2005cs,Anselmino:2005an,Anselmino:2008sga,Arnold:2008ap}.     
All of them assume a flavor-independent Gaussian transverse-momentum distribution 
of the involved TMD. Although this is an oversimplification, we adopt the same choice. 
At the starting scale $Q_0$ and following the notation of Ref.~\cite{Boer:2011fh}, we use the unpolarized distribution and fragmentation functions
\begin{align}
f_1^a(x,k_{\perp}^2; Q_0^2) &= \frac{f_1^a(x; Q_0^2)}{\pi\langle k_{\perp}^2 \rangle } \, 
                          e^{-k_{\perp}^2/\langle k_{\perp}^2 \rangle}  , 
\\
D_1^a(z,P_{\perp}^2; Q_0^2) &= \frac{D_1^a(z; Q_0^2)}{\pi  \langle P_{\perp}^2 \rangle }  \, 
                          e^{-P_{\perp}^2/\langle P_{\perp}^2 \rangle}  ,
\label{e:unpolTMD}
\end{align} 
where $z$ is the fraction of the energy of the fragmenting parton $a$ carried by the detected hadron. For $f_1^a (x)$ we use the MSTW08LO set~\cite{Martin:2009iq}, for $D_1^a (z)$ we use the DSS LO set~\cite{deFlorian:2007aj}. We fix the width of the transverse-momentum distributions for the initial parton and final hadron, respectively, as
\begin{align}  
 \langle k_{\perp}^2 \rangle &= 0.14 \text{ GeV}^2 ,
&
\langle P_{\perp}^2 \rangle &= 0.42 \, z^{0.54} (1-z)^{0.37} \text{ GeV}^2 .
\label{e:avpT}
\end{align}   
These parameters have been implemented in the HERMES {\tt gmc\_trans} Monte Carlo generator and are known to give a good description of the HERMES data~\cite{Schnell:2007}. In principle, these functions 
should be evolved according to TMD evolution~\cite{Aybat:2011zv}. However, we choose here to implement only the evolution of their collinear part. 

Neglecting the contribution of heavier $c, b, t$ flavors, we parametrize the Sivers function in the following way (inspired by~\cite{Anselmino:2008sga}):
\begin{align}
\begin{split}
f_{1T}^{\perp a}(x,k_{\perp}^2;Q_0^2) &=  f_{1T}^{\perp (0) a}(x; Q_0^2) 
\\
& \quad \frac{M_1^2 + \langle k_{\perp}^2 \rangle}{\pi M_1^2 \langle k_{\perp}^2 \rangle }\, 
         e^{-k_{\perp}^2/M_1^2}  e^{-k_{\perp}^2/\langle k_{\perp}^2 \rangle}
\end{split}
\label{e:gauss}
\end{align}
where $M_1$ is a free parameter related to the width of the transverse-momentum distribution, and 
\begin{align} 
\begin{split} 
f_{1T}^{\perp (0) q_v}(x; Q_0^2) &= C^{q_v} \sqrt{2 e} \
       \frac{M M_1}{M_1^2 + \langle  k_{\perp}^2 \rangle}\,
\\  
&\quad \frac{1- x / \alpha^{q_v}}{|\alpha^{q_v}-1|}\,(1-x) f_1^{q_v}(x; Q_0^2), 
\end{split} 
\\
f_{1T}^{\perp (0) \bar{q}}(x;Q_0^2) &= C^{\bar q}\sqrt{2 e}\
       \frac{M M_1}{M_1^2 + \langle  k_{\perp}^2 \rangle}\,  
                             (1-x) \, f_1^{\bar{q}}(x;Q_0^2) . 
\label{e:param}
\end{align}  
Note that at $Q_0$ we establish a relation between the Sivers function for the combinations $q_v$, $\bar{q}$, and the corresponding unpolarized PDF, at variance with what has been done in the
literature~\cite{Anselmino:2008sga,Arnold:2008ap}. This will turn out to be important when establishing a relation with the anomalous magnetic moment, since it guarantees that the valence Sivers function is integrable at any scale.  We multiply the unpolarized PDF by $(1-x)$ to respect the predicted high-$x$ behavior of the Sivers function~\cite{Brodsky:2006hj}. We introduce the free
parameter  $\alpha^{q_v}$ to allow for the presence of a node in the Sivers function at 
$x=\alpha^{q_v}$, as suggested by diquark model 
calculations~\cite{Bacchetta:2008af,Bacchetta:2010si}  and phenomenological 
studies~\cite{Kang:2011} (see the discussion in Ref.~\cite{Boer:2011fx}).  We imposed constraints on the parameters $C^a$ in order to respect the positivity bound for the Sivers 
function~\cite{Bacchetta:1999kz}, neglecting the contribution of the helicity distribution $g_1(x)$ (as in Ref.~\cite{Anselmino:2008sga}). For the gluons, we assume the same functional dependence of the sea quarks, Eq.~\eqref{e:param}, with the replacement $\bar{q} \to g$. 

Also for $f_{1T}^{\perp}$, we neglect the effect of TMD scale evolution~\cite{Kang:2011mr}. We assume that $f_{1T}^{\perp (0)}(x; Q^2)$ evolves in the same way as $f_1(x; Q^2)$, based on the results of Refs.~\cite{Kang:2008ey,Vogelsang:2009pj} (note however that a slightly different result has been obtained in Ref.~\cite{Braun:2009mi}). 

In conclusion, we describe the SIDIS Sivers asymmetry in the following way:
\begin{equation} 
\begin{split} 
& A_{UT}^{\sin(\phi_h-\phi_S)} (x,z,P_{T}^2,Q^2) =
 - \frac{M_1^2 (M_1^2 + \langle k_{\perp}^2 \rangle)}
           {\langle P_{\mathrm{Siv}}^2\rangle^2} \,  \frac{z\  P_{T}}{M} 
\\
&  \quad \biggl( z^2 + \frac{\langle P_{\perp}^2\rangle}{\langle k_{\perp}^2 \rangle}\biggr)^3  
    e^{- \frac{z^2 P_{T}^2}{\langle P_{\mathrm{Siv}}^2 \rangle}}
   \frac{\sum_a e_a^2 \  f_{1T}^{\perp (0) a}(x; Q^2)\ D_1^a (z;Q^2)}
           {\sum_a e_a^2 \  f_{1}^{a}(x;Q^2)\ D_1^a (z;Q^2)},
\end{split} 
\label{e:asym}
\end{equation} 
where
\begin{equation} 
\langle P_{\mathrm{Siv}}^2 \rangle = M_1^2 
     \biggl( z^2 + \frac{\langle P_{\perp}^2 \rangle}{\langle k_{\perp}^2 \rangle} \biggr) 
     \biggl( z^2 +\frac{\langle P_{\perp}^2 \rangle}{\langle k_{\perp}^2 \rangle} +
                 \frac{\langle P_{\perp}^2 \rangle}{M_1^2 } \biggr) , 
\label{e:avPhperp}
\end{equation} 
and $P_{T}$ is the modulus of the transverse momentum of the detected final hadron in the lab frame. 

For the lensing function we use the following Ansatz
\begin{equation}
L(x) = \frac{K}{(1-x)^{\eta}}.
\label{e:lensing}
\end{equation} 
The choice of this form is guided by model
calculations~\cite{Burkardt:2003je,Lu:2006kt,Meissner:2007rx,Bacchetta:2008af,Bacchetta:2010si}, 
by the large-$x$ limit of the GPD $E$~\cite{Brodsky:2006hj}, and by the phenomenological analysis of the GPD $E$ proposed in Ref.~\cite{Guidal:2004nd}. We checked {\it a posteriori} that there is no violation of the positivity bound on the GPD $E^{q_v}$ as expressed in 
Ref.~\cite{Diehl:2004cx}, again neglecting the contribution of $g_1(x)$. The nucleon anomalous magnetic moments are computed as 
\begin{align}
\kappa^p &= \int_0^1 \frac{d x}{3}  \biggl[2 E^{u_v}(x,0,0) -E^{d_v}(x,0,0)  -E^{s_v}(x,0,0)\biggr],
\nonumber \\
\kappa^n &= \int_0^1 \frac{d x}{3}  \biggl[2 E^{d_v}(x,0,0) -E^{u_v}(x,0,0)  -E^{s_v}(x,0,0)\biggr].
\label{e:kappa}
\end{align} 

We perform a combined $\chi^2$ fit to 105 HERMES proton data~\cite{Airapetian:2009ti}, to 104 COMPASS deuteron data~\cite{Alekseev:2008dn}, and to 8 JLab neutron data~\cite{Qian:2011py},
of the Sivers asymmetry with identified hadrons. We sum the statistical and systematic errors in quadrature and neglect the experimental normalization uncertainty. Since the HERMES and COMPASS data are presented as three projections of the same data set (binned in three different ways: in $x$, $z$, $P_{h \perp}$), we consider all three projections but we multiply their
statistical errors by a factor  $\sqrt{3}$ and we divide by $3$ the number of these bins (105 and 104) when counting the number of degrees of freedom. The anomalous magnetic moments are
known to a precision of $10^{-7}$ or higher~\cite{Nakamura:2010zzi}.  However, given the typical uncertainties on PDF extractions, our computation of $\kappa$ is affected by a theoretical error of the order of $10^{-3}$. Therefore,  for our present purposes we take 
$\kappa^p = 1.793 \pm 0.001, \; \kappa^n = -1.913 \pm 0.001$.

We started from considering 15 free parameters. They are $C^{\bar{q}}, \ C^{q_v}, \alpha^{q_v}$, with $q=u, d, s$, the gluon coefficient $C^g$, $M_1$, the lensing parameters $K$ and $\eta$, and the scales $Q_0$ and $Q_L$.  However, after some explorations, we made a common set of assumptions in all attempted fits. In all cases, we fixed $\alpha^{d_v,s_v}=0$ (no nodes in the valence down and strange Sivers functions, as suggested in
Refs.~\cite{Bacchetta:2008af,Bacchetta:2010si,Kang:2011,Boer:2011fx}). We also set $C^g=0$ (the influence of the gluon Sivers function through evolution is anyway limited). Finally, all fits indicated that  $Q_0 = Q_L=1$ GeV was an acceptable choice. Therefore, the actual number of free parameters is at most 10. In this framework, we conclude that it is possible to give a simultaneous description of the SIDIS data and of the nucleon anomalous magnetic moments assuming the relation in Eq.~\eqref{e:EtoSivers0}. 

We explored several scenarios characterized by different choices of the parameters related to the strange quark. We considered fits with fixed $C^{\bar{s}}=0$, or with fixed $C^{s_v}=0$, or with both parameters free (but constrained within positivity limits),  or with both fixed
$C^{s_v} = C^{\bar{s}} =0$. In all cases, we obtained very good values of $\chi^2$ per degree of freedom ($\chi^2$/dof) between 1.323 and 1.347. All fits lead to a negative Sivers function for 
$u_v$ and large and positive for $d_v$, in agreement with previous  
studies~\cite{Vogelsang:2005cs,Anselmino:2005an,Anselmino:2008sga,Arnold:2008ap}
and with some models~\cite{Courtoy:2008vi,Courtoy:2008dn,Pasquini:2010af}. The data are compatible with vanishing sea-quark contributions (with large uncertainties). However, in the $x$ range where data exist, large Sivers functions for $\bar{u}$ and $\bar{d}$ are excluded, as well as large and negative for $\bar{s}$. The Sivers function for $s_v$ is essentially unconstrained. The parameter $M_1$ is quite stable around 0.34 GeV, as well as the strength of the lensing function $K$ around 1.86 GeV. The parameter $\eta$ is typically around 0.4 but can vary between 0.03 and 2. The node $\alpha^{u_v}$ appears only above  $x \approx 0.78$.   

\begin{table}
\begin{tabular}{c|c|c|c}
\hline\hline\\[-4.5mm]
$C^{u_v}$ & $C^{d_v}$ & $C^{\bar{u}}$ & $C^{\bar{d}}$ \\[0.5mm]
\hline
$-0.229 \pm 0.002$  &  $1.591\pm 0.009$  &  $0.054 \pm 0.107$  &  $-0.083 \pm 0.122$  \\
\hline\hline
$M_1$ [GeV] & $K$ [GeV] & $\eta$ & $\alpha^{u_v}$  \\[0.5mm]
\hline
$0.346 \pm 0.015$  &  $1.888 \pm 0.009$  &  $0.392 \pm 0.040$  &  $0.783 \pm 0.001$  \\
\hline\hline
\end{tabular}
\caption{Best-fit values of the 8 free parameters for the case $C^{s_v} = C^{\bar{s}} = 0$. 
The 
final $\chi^2$/dof is 1.323. The errors are statistical and correspond to $\Delta \chi^2 = 1$} 
\label{t:parms}
\end{table}
We now discuss in detail the case with fixed $C^{s_v} = C^{\bar{s}} =0$, because it gives the best  
$\chi^2$/dof  (1.323) and suggests that it is possible to fit the present SIDIS data for Sivers asymmetries in kaon emission without the strange contribution to the Sivers function. The best-fit values of the parameters are listed in Tab.~\ref{t:parms} together with their statistical errors corresponding to $\Delta \chi^2 = 1$.  

\begin{figure}
\begin{center}
\includegraphics[width=0.3\textwidth]{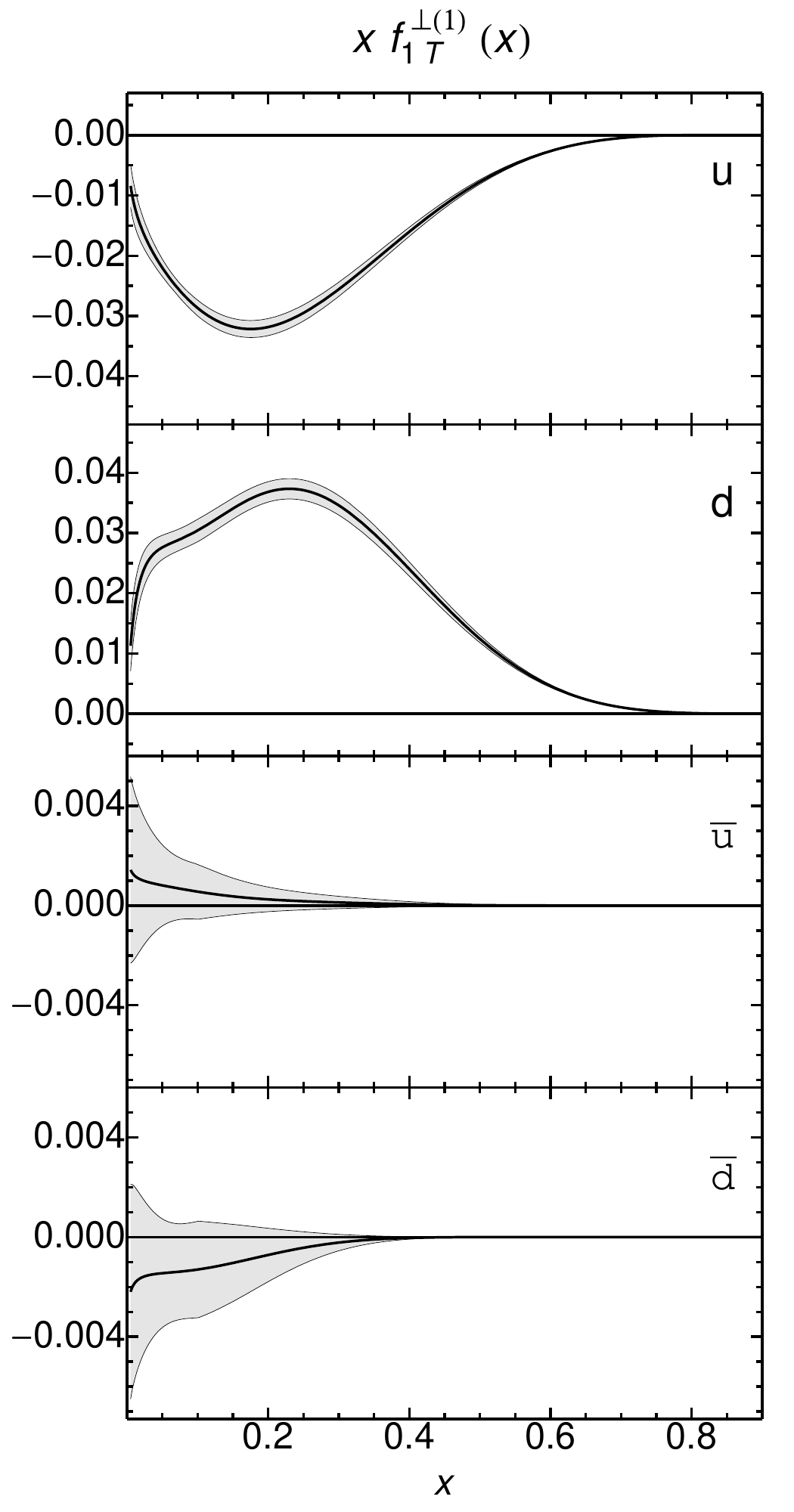}
\end{center}
\caption{\label{f:sivers} The function $x f_{1T}^{\perp (1) a} (x; Q_0^2)$ (see text) as a function of 
$x$ at the scale $Q_0 = 1$ GeV for $a = u, d, \bar{u}, \bar{d}$ from top panel to bottom, respectively. The uncertainty bands are produced by the statistical errors on the fit parameters listed in Tab.~~\ref{t:parms}.} 
\end{figure} 
In Fig.~\ref{f:sivers}, we show the corresponding outcome for $x f_{1T}^{\perp (1) a} (x; Q_0^2)$ with $a = u, d, \bar{u}, \bar{d}$. The Sivers functions for $s,\  \bar{s}$ vanish identically. The uncertainty bands are produced by propagation of the statistical errors of the fit parameters including their correlations, and correspond to  $\Delta \chi^2 = 1$. Our results are comparable with other extractions of the Sivers 
function~\cite{Vogelsang:2005cs,Anselmino:2008sga,Arnold:2008ap}. They are also qualitatively similar to the forward limit of the GPD $E$ extracted from 
experiments~\cite{Diehl:2004cx,Guidal:2004nd,Ahmad:2006gn,Goloskokov:2008ib}. 
 
We can now compute the contribution to the anomalous magnetic moment of each valence quark flavor $q_v$ using Eqs.~\eqref{e:kappa}. We obtain
\begin{align*}
\kappa^{u_v} &= 1.673 \pm 0.003^{+0.011}_{-0.000},
&
\kappa^{d_v} &= -2.033 \pm 0.002^{+ 0.011}_{-0.000},  
\\
\kappa^{s_v} &= 0^{+0.011}_{-0.000}  . 
\end{align*} 
The first symmetric error is statistical and comes again from the errors of the fit parameters 
($\Delta \chi^2 = 1$). The second asymmetric error is purely theoretical. It is computed by considering the other possible scenarios (corresponding to different choices for $C^{s_v}$ and 
$C^{\bar{s}}$) which give good $\chi^2$ fits as well. However, a precise estimate of this error can  be obtained only by performing a neural network fit~\cite{Ball:2010de}. The strange contribution to the anomalous magnetic moment is negligible, because the positivity bounds severely limit the Sivers function for $s$ and, in turn, also $E^{s_v}$ and $\kappa^{s_v}$. Our results are similar to other estimates of the strange Pauli form factor~\cite{Young:2007zs,Diehl:2007uc} and lattice QCD 
calculations~\cite{Wang:1900ta,Hagler:2011zz}.  

Using Eq.~\eqref{e:Jdef}, we can compute the total longitudinal angular momentum carried by each flavor $q$ and $\bar{q}$ at our initial scale $Q_L^2 = 1$ GeV$^2$. Using the standard evolution equations for the angular momentum (at leading order, with 3 flavors only, and $\Lambda_{\mathrm{QCD}} = 257$ MeV), we obtain the following results at $Q^2 = 4$ GeV$^2$:  
\begin{align*} 
J^u &= 0.229 \pm 0.002^{+0.008}_{-0.012} ,
&
J^{\bar{u}} &= 0.015 \pm 0.003^{+0.001}_{-0.000} ,  
\\
J^d &= -0.007 \pm 0.003^{+0.020}_{-0.005} ,
&
J^{\bar{d}} &= 0.022 \pm 0.005^{+0.001}_{-0.000} ,  
\\
J^s &= 0.006^{+0.002}_{-0.006} ,
&
J^{\bar{s}} &= 0.006^{+0.000}_{-0.005} . 
\end{align*} 
As before, the first symmetric error is statistical and related to the errors on the fit parameters, while the second asymmetric error is theoretical and reflects the uncertainty introduced by the other possible scenarios. In the present approach, we cannot include the (probably large) systematic error due to the rigidity of the functional form in Eqs.~\eqref{e:gauss}-\eqref{e:param},~\eqref{e:lensing}. The bias induced by the choice of the functional form may affect in particular the determination of the sea quark angular momenta, since they are not directly constrained by the values of the nucleon anomalous magnetic moments. Our present estimates (at $Q^2 = 4$ GeV$^2$) agree well with other analyses~\cite{Diehl:2004cx,Guidal:2004nd,Ahmad:2006gn,Goloskokov:2008ib,Wakamatsu:2007ar,Bratt:2010jn}. It indicates a total contribution to the nucleon spin from quarks and antiquarks of $0.271 \pm 0.007^{+0.032}_{-0.028}$, of which 85\% is carried by the up quark. 

In summary, we have presented a determination of the quark angular momentum assuming a connection between the collinear limit of the generalized parton distribution $E$ and the Sivers transverse-momentum distribution. We have shown that it is possible to fit at the same time the nucleon anomalous magnetic moments and data for semi-inclusive single-spin asymmetries produced by the Sivers effect. Several different scenarios produce equally good $\chi^2$ fits.  
Our strategy opens a plausible way to quantifying the quark angular momentum, and imposes additional constraints on the Sivers function.  

We thank B. Pasquini for critically reading the manuscript. We thank the Institute for Nuclear Theory at the University of Washington and Brookhaven National Laboratory for their hospitality, and the Department of Energy for partial support during the completion of this work. This work is partially supported by the Italian MIUR through the PRIN 2008EKLACK, and by the European Community through the Research Infrastructure Integrating Activity ``HadronPhysics2" (Grant Agreement n. 227431) under the $7^{\mathrm{th}}$ Framework Programme.


\bibliographystyle{h-physrev}
\bibliography{mybiblio}


\end{document}